# Comb mode converters based on sampled helical gratings

Yancheng Ma[1], Liang Fang[2] and Guoan Wu[*1]

[1] School of Optical and Electronic Information, Huazhong University of Science and Technology, Wuhan 430074, Hubei, China

[2] Wuhan National Laboratory for Optoelectronics, School of Optical and Electronic Information, Huazhong University of Science and Technology, Wuhan 430074, Hubei, China

E-mail: GuoanWu_HUST@163.com

Abstract: Helical gratings (HGs) have an achievement of flexible mode conversion for fiber-guided orbital angular momentum (OAM) modes. Sampled reflection HGs can realize the generation and conversion of OAM mode with comb spectra. They can be used to simultaneously filter wavelengths and convert modes, and thus may be applied to the hybrid multiplexing technology with wavelength division multiplexing (WDM) and mode division multiplexing (MDM).

## 1. Introduction

In recent years, with the ever-increasing demand for fiber bandwidth in global information, single-mode fiber (SMF), a finite bandwidth capacity, is insufficient to satisfy the requirements [1]. To expand the bandwidth capacity, after the wavelength division multiplexing (WDM) technology [2], space-division multiplexing (SDM) [3], mode-division multiplexing (MDM) [4-6] including orbital angular momentum (OAM) multiplexing [7] as new fiber technologies have recently been proposed and achieved experimentally. It has been well known that OAM as the new degree of freedom of light has the great prospect of broadening the optical bandwidth [8]. As for fiber-guided OAM modes, vortex fiber or ring-core fiber are usually used to support them with the lowest radial order, and have the characteristic of the large difference of effective refractive index ($\Delta n_{eff} > 10^{-4}$) within the vector mode groups that may tend to degenerate into linearly polarized (LP) mode in general fibers [9, 10]. Therefore, these fibers are available to transmit OAM modes with low inter-mode crosstalk in optical communication with fiber-guided OAM modes. It has been theoretically studied that HGs inscribed in ring-core fibers can realize flexible generation, conversion, and exchange of fiber-guided OAM modes, which belongs to all-fiber method of converting OAM modes in fibers [11]. In this Letter, we develop the theoretical study for HGs, sampling the modulation of HGs in a ring-core fiber to achieve OAM mode conversion with comb spectra, analogous to conventional sampled fiber Bragg gratings (FBGs) in SMF, but they just can reflect the fundamental mode with comb spectra. The function of converting OAM modes with comb spectra by sampled HGs can simultaneously filter wavelengths and convert OAM modes carried by these wavelengths, and thus may be applied to operate wavelengths and modes in the hybrid multiplexing technology with both WDM and MDM to improve the channel performance of the current optical communication system [12].

## 2. Principle and simulation

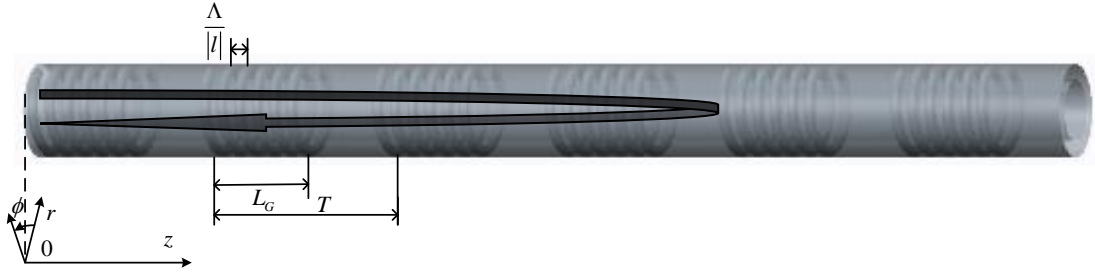

**Fig. 1** Schematic diagram of sampled HGs $\langle -1,1 \rangle$ inscribed in a ring-core fiber

The sampled HGs inscribed in ring-core fiber can be formed by rotating and moving the fiber when writing them by the means of point-by-point or a phase mask [13, 14]. The modulation function of a sampled $|l|$-fold HGs denoted with $\langle \sigma, |l| \rangle$ can be expressed by

$$\delta n_s(r,\phi,z) = \Delta n \cdot \rho(r) \cdot \sum_{i=0}^{N} rect\left(\frac{z - i \cdot T}{L_G}\right) \cdot \cos\left[l\left(\phi + \sigma \frac{2\pi}{\Lambda} z\right)\right] \quad (a_1 \leq r \leq a_2),$$

where $\Delta n$ is the modulation strength, the absolute value of $l$ indicates the fold number of helical fringes of sampled HGs, $\Lambda$ denotes their period, $T$ sampling period, $N$ the number of sampling modulation, $L_G$ the grating length of each HGs element, $\sigma$ the helix orientation, $\sigma = +1$ and $-1$ correspond to left-handed and right-handed helix, respectively, $\rho(r)$ determinates the saturability of index modulation in the cross section defined by $\rho(r) = r^2/a_2^2$, $a_1$ and $a_2$ correspond to inside and outside radii of the structure of ring-core waveguide, respectively. Fig. 1 shows the structure and modulation fringes of sampled HGs $\langle -1,1 \rangle$, where the arrow represents the reflection of coupled modes.

The polarization dependent OAM mode is denoted with $|s_n, n\rangle$ where $n$ and $s_n$ denote the topological charge of OAM and SAM, respectively, and $s_n = -1$, $+1$ or $0$ correspond to right circular, left circular or linear polarization, respectively. The amplitude conversion between the transmission modes $|s_n, n\rangle$ with amplitude $A_i$ and reflection modes $|s_m, m\rangle$ with amplitude $B_i$ through the $i$-th grating section can be written

$$\begin{bmatrix} A_i \\ B_i \end{bmatrix} = \mathbf{F}_i^G \cdot \begin{bmatrix} A_{i-1} \\ B_{i-1} \end{bmatrix}$$

where transfer matrix

$$\mathbf{F}_i^G = \begin{bmatrix} \cosh(\gamma \Delta z) - i\frac{\hat{\sigma}}{\gamma}\sinh(\gamma \Delta z) & -i\frac{\kappa_{nm}}{\gamma}\sinh(\gamma \Delta z) \\ i\frac{\kappa_{nm}}{\gamma}\sinh(\gamma \Delta z) & \cosh(\gamma \Delta z) + i\frac{\hat{\sigma}}{\gamma}\sinh(\gamma \Delta z) \end{bmatrix}$$

where

$$\hat{\sigma} = -\frac{1}{2}\left(\beta_m + \beta_n + \sigma l \frac{2\pi}{\Lambda}\right)$$

with the propagation constant of OAM modes $|s_n,n\rangle$ and $|s_m,m\rangle$ defined $\beta_n = 2\pi n_{eff1}/\lambda$ and $\beta_m = 2\pi n_{eff2}/\lambda$, and $n_{eff1}$, $n_{eff2}$ corresponding to their effective index, respectively, and

$$\gamma = \sqrt{\kappa_{nm}^2 - \hat{\sigma}^2}$$

with coupling coefficient for reflection HGs

$$\kappa_{nm} = \frac{\omega\varepsilon_0 n_2^2 \Delta n(1-s_n s_m)}{4a_2^2}\int_{a_1}^{a_2}\int_0^{2\pi} F_n(r)^* \cdot F_m(r)\exp\left[i(m+n-l)\phi\right]r^3 drd\phi$$

where $*$ denotes operation of conjugate transpose, $\omega$ and $\varepsilon_0$ are angular frequency and dielectric constant in vacuum, respectively, and $n_2$ is the refractive index of the ring-core waveguide, $F_n(r)$ and $F_m(r)$ radial function of electric field of corresponding OAM modes. The detailed coupling rule of reflection HGs is revealed in reference [11]. One can find the matrix elements $F_i^G(1,2)$ and $F_i^G(2,1)$ of the $i$-th grating section are responsible for amplitude conversion between two OAM modes carrying different OAM. The transfer matrix of duty section of sampled HGs can be written

$$\mathbf{F}_i^D = \begin{bmatrix} \exp(-ik_0 n_{eff1} L_D) & 0 \\ 0 & \exp(ik_0 n_{eff2} L_D) \end{bmatrix}$$

then transfer matrix of one sampling period

$$\mathbf{F}_i = \mathbf{F}_i^G \cdot \mathbf{F}_i^D.$$

Finally, the transfer matrix of total sampled HGs can be obtained

$$\mathbf{F} = \mathbf{F}_N \cdot \mathbf{F}_{N-1} \cdot ... \cdot \mathbf{F}_i \cdot ... \cdot \mathbf{F}_2 \cdot \mathbf{F}_1$$

and the ultimate amplitude transferring through sampled HGs is

$$\begin{bmatrix} A(L) \\ B(L) \end{bmatrix} = \mathbf{F} \cdot \begin{bmatrix} A(0) \\ B(0) \end{bmatrix}.$$

When the boundary conditions here is $A(0)=1$ and $B(L)=0$, one can obtain $A(L)$ and $B(0)$, and thus the reflectivity of sampled HGs defined is $r = |B(0)/A(0)|^2$.

In our simulation, the total length of sampled HGs is $L=4.0$ cm, the sampling period is $T=2.0$ mm, and the duty ratio is $D=0.18$, and thus the grating length of each HGs element is $L_G = D \cdot T = 0.36$ mm, the modulation strength is taken $\Delta n = 3\times 10^{-4}$. In Fig. 2, we get the reflection spectrum of OAM mode $|+1,+1\rangle$ when the incident mode is the circular fundamental mode $|-1,0\rangle$ using the sampled HGs $\langle-1,1\rangle$ with grating period $\Lambda = 467.0$ nm. One can see that an equidistance 8 channel spectrum is obtained nearby the wavelength of 1550nm, and meanwhile each channel carries OAM with topological charge +1. Note that this comb spectrum is different from the conventional sampled FBGs that just reflect the fundamental mode with comb spectrum[15].

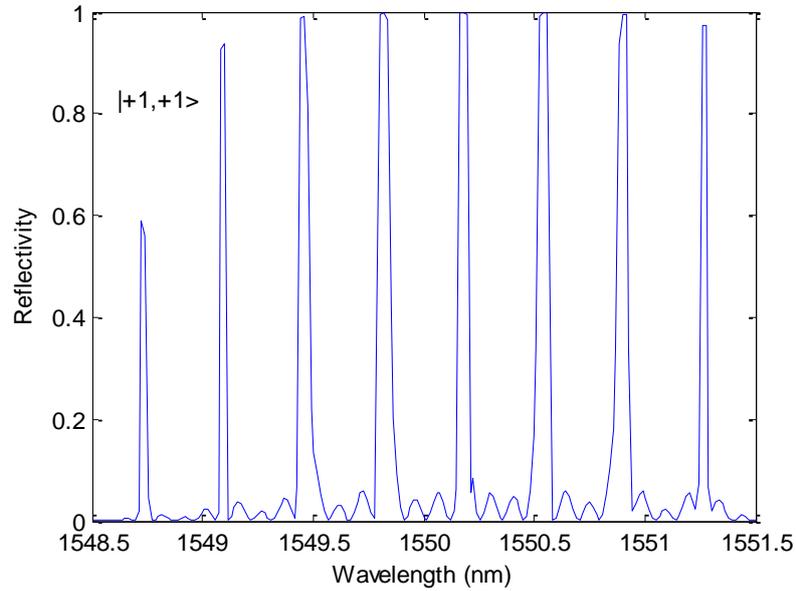

**Fig. 2** Reflective spectrum of OAM modes $|+1,+1\rangle$ by sampled HGs $\langle-1,1|$, the incident modes is $|-1,0\rangle$.

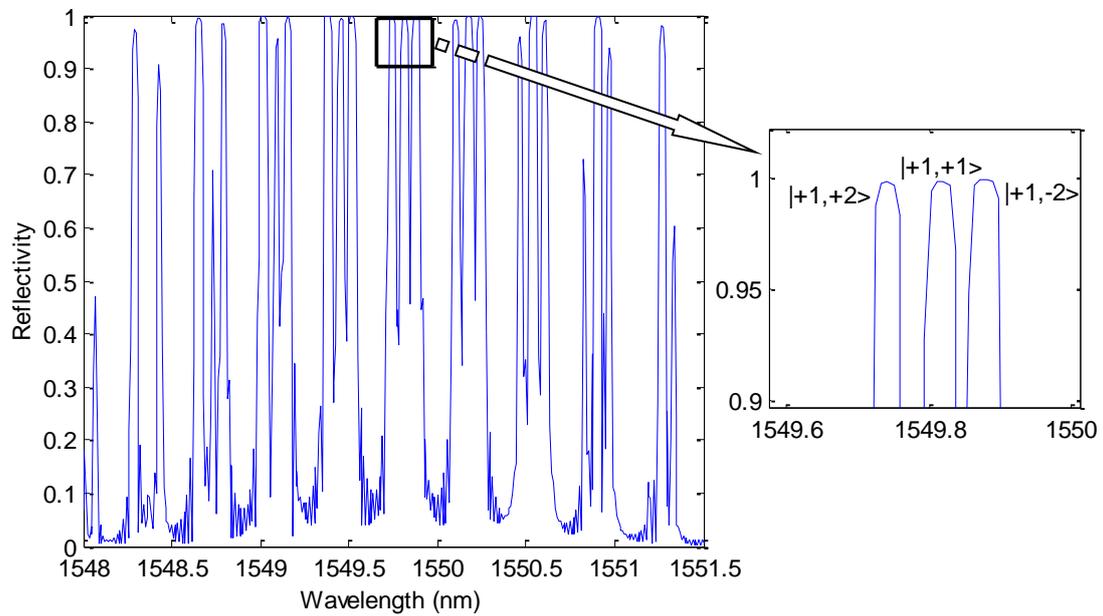

**Fig. 3** Reflective spectrum of OAM modes $|+1,+1\rangle$, $|+1,+2\rangle$ and $|+1,-2\rangle$ by three corresponding sampled HGs $\langle-1,1|$, $\langle-1,2|$ and $\langle+1,2|$, respectively, cascaded in one ring-core fiber, the incident modes is $|-1,0\rangle$.

When several sampled HGs are successively cascaded in one ring-core fiber, we can get a hybrid comb spectrum carrying different OAM. We simulate the three cascaded sampled HGs, and get the comb spectrum with multi-wavelength and three OAM states shown in Fig. (3). The sampled HGs are taken as $\langle -1,1 \rangle$, $\langle -1,2 \rangle$ and $\langle +1,2 \rangle$ with corresponding grating period $\Lambda = 467.0\,\text{nm}$, $934.9\,\text{nm}$ and $934.7\,\text{nm}$, respectively. The magnified figure shows that each group of peaks at the comb spectrum consists of three kinds of peaks carrying left circularly polarized OAM states with topological charges +2, +1 and -2, respectively, from left to right. The interval between these OAM peaks is very sensitive to the grating period of each of the sampled HGs. The sampled HGs exhibit the coupling manner in simultaneously filtering wavelengths and converting OAM modes carried by these wavelengths. It may be meaningful to design this type of sampled HGs to generate the alike comb spectra carrying hybrid modes or to operate wavelengths and modes in a hybrid multiplexing technology with both WDM and MDM. It is likely that this grating structure can decrease the crosstalk relying on the orthogonality between different OAM modes and increase the bandwidth capacity of multiplexed channels for this hybrid multiplexing technology .

3. **Summary**

In this Letter, we have theoretically analyzed the coupling features of the sampled HGs based on the coupled-mode theory and the methods of transfer matrix. The sampled HGs can generate the comb spectrum carrying OAM dependent of the fold number, the grating period and helix orientation of HGs. When cascading several sampled HGs in one ring-core fiber, the hybrid comb spectrum carrying different OAM can be achieved. This special coupling manner of sampled HGs may be potentially to applied to the hybrid multiplexing technology with WDM and MDM to improve the channel performance of the current optical communication system.

Reference

[1] Richardson, D. J., Fini, J. M. and Nelson, L. E., "Space-division multiplexing in optical fibers", Nature Photonics, 2013, **7**, (5), pp. 354-362

[2] Shimada,T., Sakurai, N., and Kumozaki, K., "WDM access system based on shared demultiplexer and MMF links", J. Lightwave Technol., 2005, **23**, (5), pp. 2621-2628

[3] Winzer, P. J., "Making spatial multiplexing a reality", Nature Photonics, 2012, **8**, (5), pp. 345–348

[4] Koebele, C., Salsi, M., Sperti, D., et al., "Two mode transmission at 2×100Gb/s, over 40km-long prototype few-mode fiber, using LCOS-based programmable mode multiplexer and demultiplexer", Opt. Express, 2011, **19**, (17), pp. 16593-16600

[5] Fang, L., and Jia, H., "Mode add/drop multiplexers of $LP_{02}$ and $LP_{03}$ modes with

two parallel combinative long-period fiber gratings", Opt. Express, 2014, **22**, (10), pp. 11488-11497

[6] Fang, L., and Jia, H., "Coupling analyses of LP0m modes withoptical fiber gratings in multimode fiber and their application in mode-division multiplexing transmission", Opt. Commun., 2014, **322**, pp. 118-122

[7] Bozinovic, N., Yue, Y., Ren, Y. X., et al., "Terabit-scale orbital angular momentum mode division multiplexing in fibers". Science, 2013, **340,** pp. 1545-1548

[8] Wang, J., Yang, J., Faza, I. M., et al., "Terabit free-space data transmission employing orbital angular momentum multiplexing", Nature Photonics, 2012, **22**, (6), pp. 488–496

[9] Ramachandran, S., Kristensen, P., and Yan, M. F., "Generation and propagation of radially polarized beams in optical fibers", Opt. Lett., 2009, **34**, (16), pp. 2525-2527

[10] Ramachandran, S., and Kristensen, P., "Optical vortices in fiber", Nanophotonics, 2013, **2**, (5-6), pp. 455–474

[11] Fang, L., and Wang, J., "Flexible generation/conversion/exchange of fiber-guided orbital angular momentum modes using helical gratings", Opt. Lett., 2015, **40**, (17), pp. 4010-4013

[12] Fazal, I. M., Ahmed, N., Wang, J., et,al., "2 Tbit/s free-space data transmission on two orthogonal orbital-angular-momentum beams each carrying 25 WDM channels", Opt. Lett., 2012, **37**, (22), pp. 4753-4755

[13] Lin, Z., Wang, A., Xu, L., et,al., "Generation of Optical Vortices Using a Helical Fiber Bragg Grating" J. Lightwave Technol., 2014, **32**, (11), pp. 2152-2156

[14] Williams, R. J., Krämer, R. G., Nolte, S., and Withford, M. J., "Femtosecond direct-writing of low-loss fiber Bragg gratings using a continuous core-scanning technique" Opt. Lett., 2013 **38**, (11) 1918-1920.

[15] Eggleton, B. J., Krug, P. A., Poladian, L., and Ouellette, F., "Long periodic superstructure Bragg gratings in optical fibres", Electron. Lett., 1994. **30**, (19), pp. 1620-1622